# Ultra-hard carbon film from epitaxial two-layer graphene

Yang Gao[1,2*], Tengfei Cao[1,3*], Filippo Cellini[1], Claire Berger[2,4], Walt de Heer[2], Erio Tosatti[5,6], Elisa Riedo[1,2,7,8,†] and Angelo Bongiorno[3,8,9,†]

[1] Advanced Science Research Center, City University of New York, 85 St Nicholas Terrace, New York, New York 10031, USA

[2] School of Physics, Georgia Institute of Technology, 837 State Street, Atlanta, Georgia 30332, USA

[3] Department of Chemistry, College of Staten Island, City University of New York, 2800 Victory Boulevard, Staten Island, New York 10314, USA

[4] Institut Néel, CNRS- University Grenoble-Alpes, 38042 Grenoble, France

[5] Abdus Salam ICTP, Strada Costiera 11, 34151 Trieste, Italy

[6] SISSA, via Bonomea 265, 34136 Trieste, Italy

[7] Physics Department, City College of New York, City University of New York, Marshak Science Building, 160 Convent Avenue, New York, New York 10031, USA

[8] CUNY Graduate Center, Ph.D. Program in Physics, New York, NY 10016

[9] CUNY Graduate Center, Ph.D. Program in Chemistry, New York, NY 10016

* Contributed equally to this work

[†] Corresponding authors:

Email: elisa.riedo@asrc.cuny.edu
Email: angelo.bongiorno@csi.cuny.edu

**Atomically thin graphene exhibits fascinating mechanical properties, although its hardness and transverse stiffness are inferior to those of diamond. To date, there hasn't been any practical demonstration of the transformation of multi-layer graphene into diamond-like ultra-hard structures. Here we show that at room temperature and after nano-indentation, two-layer graphene on SiC(0001) exhibits a transverse stiffness and hardness comparable to diamond, resisting to perforation with a diamond indenter, and showing a reversible drop in electrical conductivity upon indentation. Density functional theory calculations suggest that upon compression, the two-layer graphene film transforms into a diamond-like film, producing both elastic deformations and $sp^2$-to-$sp^3$ chemical changes. Experiments and calculations show that this reversible phase change is not observed for a single buffer layer on SiC or graphene films thicker than 3 to 5 layers. Indeed, calculations show that whereas in two-layer graphene layer-stacking configuration controls the conformation of the diamond-like film, in a multilayer film it hinders the phase transformation.**

Carbon can form different types of materials, exhibiting a variety of exceptional properties, from diamond to graphite, and more recently fullerenes, nanotubes, and graphene. Sintering materials with a stiffness and hardness equal or superior to those of diamond is an ongoing challenge [1-3]. Thus, for both scientific and technological reasons, the transformation of graphite into diamond remains one of the most fascinating and studied solid-to-solid phase transitions in materials science [2,4-12]. Recently, experimental [13-16] and theoretical [10,17,18] efforts have been directed to study the transformation of multi-layer graphene into a diamond-like structure induced by chemical functionalization of the surface layers. Despite recent successes [13,14], the experimental proof that such atomically-thin diamond-like films exhibit mechanical properties similar to those of diamond remains, unfortunately, missing.

Here we use sub-Å resolution modulated nano-indentation (MoNI) [19] atomic force microscopy (AFM) experiments, micro-hardness and conductive AFM measurements, and density functional theory (DFT) calculations to investigate the mechanical properties of multi-layer graphene films on SiC. Our study shows that at room temperature, indentation in a two-layer epitaxial graphene film grown on SiC(0001) induces a reversible transformation from a graphitic to a diamond-like structure, resulting in stiffness and hardness values comparable to those of diamond. In particular, in our MoNI experiments [19], a Si spherical probe with a radius of 120 nm is used to indent SiC and epitaxial graphene films of various thicknesses, reaching loads of the order of 100 nN, corresponding to pressures of about 1-10 GPa considering the contact area between the spherical tip and the flat substrate, and indentation depths ranging from 0.6 to 6 Å, depending on the thickness of the multilayer graphene film. After indenting a multi-layer graphene film on SiC, force versus indentation curves are collected while retracting the tip with sub-Å resolution. These experiments show that the transverse elastic modulus of multi-layer graphene on SiC(0001) with a number of layers larger than 5 is close to the Young's modulus of graphite in the direction perpendicular to the basal plane, i.e. ~30 GPa, whereas for a single carbon layer on SiC(0001) – the graphene buffer layer – the Young's modulus is the same as that of a bare SiC substrate. Surprisingly, our MoNI measurements show that the transverse stiffness of two-layer (2-L) epitaxial graphene on SiC(0001) is much larger than that of the bare substrate, i.e. > 400 GPa. Furthermore, micro-hardness measurements show that 2-L epitaxial graphene resists plastic deformations upon indentation with a diamond indenter, even at the maximum loads of 30 µN; whereas similar loads can easily produce *holes* in both SiC and multi-layer epitaxial graphene for films thicker than 5-L. DFT calculations indicate that the unique mechanical response to indentation of 2-L graphene stems from its ability to undergo a structural transformation into a diamond-like film, and that such ability is hindered in 3-L or thicker films by unfavorable layer stacking configurations. This theory is further confirmed by the experimental observation of a reversible decrease in electrical conductivity induced by indentation in 2-L graphene, absent in 5-L graphene. In particular, our DFT calculations show that the $sp^2$-to-$sp^3$ structural changes in a 2-L graphene film on SiC(0001) are facilitated by the presence of a buckled buffer layer in contact with the incommensurable and reactive Si-terminated face of the SiC substrate, that they occur upon indentation regardless of the nature of the layer stacking configuration, and that the resulting diamond-like film does not necessitate the presence of adsorbates to passivate and stabilize the surface of the film in contact with the indenter.

**Stiffness Measurements for 0, 1, 2, 5-L graphene on SiC**

Graphene films are grown epitaxially on the SiC(0001) surface by thermal sublimation [20] [21] [see Supplementary Information (SI) and Methods section for additional details]. Sub-Å resolved nano-indentation curves are obtained by modulated nano-indentation (more details are reported in the Methods section and SI) [19]. In a MoNI experiment, a silicon tip is brought into hard contact with a graphene film. The tip indents the sample until it reaches a normal force of 50 to 300 nN. The actual measurement is taken after forming the hard contact, i.e. during the process of retracting the AFM tip from the supported film at a rate of ~0.01 nm/s (Fig. 1a). At this point, the normal force is gradually reduced, while vertically oscillating the tip-holder with a ~0.1 Å amplitude at a fixed frequency, by means of a lock-in amplifier. For each value of the normal force between tip and sample, $F_z$ (which is controlled and kept constant by the feedback loop of the AFM), a lock-in amplifier is used to measure the slope of the force vs. indentation depth curve, namely the contact stiffness $k_{cont}(F_z)$ [19]. Force vs. indentation depth curves with sub-Å resolution are then obtained by integrating the equation $dF_z = k_{cont}(F_z)\, dz_{indent}$ as follows:

$$z_{indent}(F_z) = \int \frac{dF'_z}{k_{cont}(F'_z)} \qquad (1)$$

For simplicity, and despite its limitations, here we use the Hertz model to extract elastic moduli from our experimental force vs. indentation depth curves. In previous studies, MoNI measurements and the Hertz model were used successfully to probe and characterize the elasticity of bulk materials, including very stiff materials such as SiC (Fig. 1b), as well as the radial elasticity of nanotubes and the interlayer elasticity in graphene oxide films [19,22-24]. We underline the fact that the high resolution of the indentation curves obtained from MoNI experiments allows to measure the Young's modulus of materials that are stiffer than the Si tip.

We use the MoNI technique to probe the vertical elastic modulus of 10-L, 5-L, 2L and a single graphene layer in contact with the Si-face of SiC(0001), i.e. the buffer layer. In these experiments, the maximum indentation depth ranges from 0.6 Å in the case of 2-L graphene up to 6 Å in the case of 10-L graphene, and the experimental errors on the tip normal force, $F_z$, and indentation depth are $\Delta F_z = 0.5$ nN and $\Delta z = 0.01$ nm, respectively. To probe the number of layers in the graphene films, we perform transmission electron microscopy (TEM) and scanning TEM (STEM) measurements, as well as Raman spectroscopy and Kelvin Probe Force Microscopy (KPFM) experiments. TEM images and force vs. indentation depth curves of 5-L graphene, the buffer layer, and 2-L graphene are shown side by side in Figs. 1 and 2 (see also SI). The indentation curves shown in Fig. 1 result from averaging more than 10 curves obtained from MoNI measurements of different spots on different samples. In Fig. 1b, the indentation curve of 10-L graphene [25] is very close to that obtained for bulk Highly Ordered Pyrolytic Graphite (HOPG) [19]. The elastic modulus derived by fitting the indentation curve for 10-L graphene with a Hertz model is about 36 GPa, very close to the *c*-axis Young's modulus of graphite [19,26]; the indentation curve for 10-L reaching a depth of 6 Å is reported in the Supplementary Fig. 5. Figures 1d and 1f show the indentation curves of a bare SiC substrate together with the curves of a supported 5-L graphene film and the buffer layer, respectively. Analysis of the indentation curve of SiC yields, in agreement with published values [19], a Young's modulus of 400 GPa, whereas the curves for the buffer layer and the 5-L film clearly show that coating SiC with a graphene film leads to a softening of the elastic modulus measured by the MoNI technique. Although the exact nature of the buffer layer is still under scrutiny [21], the notion that it consists of a graphene layer strongly interacting with the SiC substrate is well accepted, and thus it does not surprise that its effect on the indentation curve is detectable but mild. On the other hand, a 5-L graphene film exhibits a transverse elastic modulus comparable to that of a graphitic system [19,26], and its softening effect of the elastic modulus of the entire system, substrate plus 5-L film, is significant and almost comparable to that of a 10-L graphene film. Overall, all these elastic behaviors are expected and well-understood to arise when a rigid substrate is coated by a thin film of a softer material of increasing thickness, and indeed the indentation curves for 10-L, 5-L and buffer layer graphene on SiC follow a predictable trend. However, and very surprisingly, when we indent two layers of graphene on SiC, as shown in Fig. 2a with the corresponding TEM image (Fig. 2b), we obtain force vs. indentation curves which are much steeper than those for 5-L or 10-L graphene and even steeper than bare SiC. Such large stiffness for 2-L epitaxial graphene corresponds to a transverse Young's modulus much larger than the modulus of SiC, i.e. $\gg$ 400 GPa, indicating that a 2-L graphene film can enhance significantly the stiffness of the substrate. Since the Hertz model is appropriate for isotropic bulk materials, here a simple Hertz fit is inadequate to extract from these elastic indentation curves the exact value of the transverse Young's modulus of the isolated atomically thin 2-L film. Nevertheless, the experimental elastic indentation curves of 2-L graphene on SiC clearly indicate that important physical/chemical effects must happen in the 2-L film to explain the stiffening effect produced by coating SiC(0001) with a 2-L graphene film. Our computational modeling work (*vide infra*), in agreement with recent theoretical works [17,18] and indentation studies of suspended single layer graphene [27], suggests that the ultra-stiffening caused by 2-L graphene is due to a transformation from a layered into a diamond-like structure occurring in the 2-L carbon film upon indentation, namely when the AFM tip approaches and presses upon the supported film.

**Micro-Hardness and Conductivity Measurements Upon Indentation**

To further challenge the existence of a diamond phase induced by pressure in two-layer epitaxial graphene, we perform micro-hardness experiments (Fig. 3a-d) with a diamond indenter using loads up to 12 μN on a 2-L and a 5-L graphene film on SiC(0001), as well as on a bare SiC substrate (more details and measurements are reported in the Methods section and Supporting Fig. 6). Quite unexpectedly, although in agreement with our MoNI measurements, upon indentation and after subsequent AFM topographic imaging, no residual indent is found in 2-L epitaxial graphene on SiC, see Fig. 3a, indicating a hardness value close if not superior to that of the diamond indenter. By contrast, when indenting a bare SiC sample with the same indenter and load, we are able to identify a residual indent (diameter ~30-40 nm), Fig. 3b. Furthermore, a larger and deeper residual indent is observed when indenting with the same conditions 5-L epitaxial graphene (diameter ~ 60-70 nm) as shown in Fig. 3c. For completeness, we remark that these experiments led us to estimate a hardness [28] [29] for SiC of 20 ± 10 GPa (see Methods for details), which is close to hardness values reported in literature *[30,31]*, as well as a value of hardness for the 5-L graphene film of the order of 5 ± 2 GPa, which is similar to the hardness of HOPG [28].

To further investigate the pressure activated carbon phase, we carry out load-dependent conductive AFM (C-AFM) experiments on 2-L and 5-L graphene films on SiC, see Fig. 3e. These experiments are conducted by using a metallic AFM tip to apply pressure and simultaneously collect the electrical current flowing between the tip and a silver electrode deposited on the graphene film surface under the influence of a constant voltage bias, see Supplementary Fig. 7. Current measurements are taken while varying the tip load from 200 up to 400 nN. Figure 3e displays the current measurements for the 5-L and 2-L graphene films on SiC. In the 5-L film, the current increases for increasing the normal load, consistent with the increase in contact area and hence in the electrical contact between tip and film. Instead, for 2-L films these experiments show that although the current initially increases with load, for loads > 260 nN the current suddenly drops, suggesting that at these loads the 2-L film undergoes a phase transition to a structure with a larger $sp^3$ content and thus more resistive. C-AFM experiments on different 2-L samples (e.g. two samples shown in Fig. 3e) but same AFM tip show that the structural transition consistently occurs at a critical load of about 270 nN. We remark that the observed decrease in electrical conductivity induced by indentation in 2-L graphene is fully reversible upon decrease of load, suggestive of a spontaneous healing of the transformed nucleus taking place at room temperature. Variations in the observed absolute values of current in the different films are attributable to inhomogeneity of the film surface, quality of the silver contact, distance between measured region and silver electrode, and tunneling phenomena at the periphery of the tip-film contact.

**DFT and Indentation Simulations**

MoNI, micro-hardness, and conductive AFM experiments show that the behavior upon compression of a 2-L graphene film on SiC(0001) is inconsistent with that of a substrate coated by a carbon film with a graphitic-like structure, which should lead to a softening of the transverse elasticity as observed in the case of a 5-L or 10-L film on SiC. Our DFT calculations and classical indentation simulations based on Hookean force fields indicate that the ultra-stiffness exhibited by 2-L graphene on SiC arises from a mechano-chemical effect, namely: in addition to deformations, the mechanical load on the supported 2-L graphene film produces $sp^2$-to-$sp^3$ structural and chemical changes, ultimately leading to the formation of a stiff diamond-like film coating the SiC substrate. Calculations and results of our modeling study are discussed in full in the SI.

A graphene film interacts strongly with the Si-face of a SiC(0001) substrate. To gain insight on how this interface may influence elastic properties and structural transformations of a carbon film supported by SiC, we consider model structures of 2-, 3-, and 4-L graphene films sandwiched between simplistic representations of a SiC(0001) surface, consisting of a single SiC(0001) layer with Si atoms in contact with the bottom and top layers of the carbon films, and the C atoms saturated by H atoms. We consider multilayer graphene films with several different layer-stacking configurations, as well as supercells of increasing lateral size to mimic different interfacial structures and account for the incommensurability of the graphene buffer layer and SiC(0001). For each model structure, we carry out a sequence of DFT optimization calculations, in which the C atoms of the SiC layers are constrained to lie on planes at a fixed separation. These DFT calculations show that a 2-L graphene can transforms upon compression into a diamond-like film in contact with SiC(0001) (Fig. 4a). The $sp^2$-to-$sp^3$ structural and chemical changes occur regardless of the stacking configuration of the two graphene layers, they are favored by the buckling distortions of the buffer layer in contact with the reactive SiC(0001) substrate, and they occur more favorably (i.e. the energy cost associated to the homogeneous transformation vanishes) by reducing the lattice mismatch at the interface between substrate and buffer layer (see Supplementary Fig. 12). Moreover, DFT calculations show that the phase transformation to a diamond-like film in compressed 3-L and 4-L graphene films is hindered by the layer-stacking configuration. In particular, our calculations show that whereas a 3- or 4-L graphene film with the ABC (rhombohedral) or AAA (hexagonal) stacking configuration can transform upon compression into a diamond-like film, a film with the AB (Bernal) stacking configuration typical of multilayer graphene and graphite undertakes only partial $sp^2$-to-$sp^3$ structural changes (Supplementary Fig. S13). In this latter case, the unfavorable layer stacking prevents the graphitic-to-diamond transition from propagating across the entire transverse direction of the film. DFT calculations based on these types of model structures (see Fig. 4a and SI) show also that a diamond-like film forms a hard contact with the SiC(0001) substrate, exhibiting an interfacial layer of bonds with a transverse elasticity (or elastic modulus) as stiff as that of the substrate material. This important result is accounted for in our indentation simulations based on atomistic Hookean force fields (*vide infra*).

We use DFT calculations to investigate the structure, mechanical, and electronic properties of diamond-like films resulting by compression of a 2-L graphene film (see Supplementary Figs. 8-10). These calculations show that, regardless of the stacking configuration, two graphene layers buckle to form a diamond-like film exposing dangling bonds towards the vacuum region. Here, we construct six possible configurations for a diamond-like film resulting by compressing a 2-L graphene film with a AA, AB, or $A_BC$ stacking configuration ($A_BC$ is here used to indicate the stacking configuration resulting when the midpoints between bonded A and B sites of one layer are aligned along the transverse direction with the C sites of the second layer). In this case, we use H to saturate the dangling bonds exposed by the surfaces of the films (Fig. 4b), and we use DFT calculations to estimate the transverse elastic modulus of the hydrogenated diamond-like films, finding values ranging from 0.30 up to 1.01 TPa, close to the Young's modulus calculated for bulk diamond. The stiffest film has the structure of hexagonal diamond (lonsdaleite [32], Supplementary Fig. 9) and it results by compressing two graphene layers stacked in an AA configuration. We also remark that two of these six conformations for a diamond-like film exhibit surface geometries showing the occurrence of broken resonance structures. DFT calculations show that the removal of H atoms from one of the two equivalent surfaces of these two diamond-like films leads to energetically stable and insulating carbon films, exhibiting an electronic band gap of about 2 eV (Fig. 4c) and exposing a clean and chemically inert surface to the vacuum. These results are in agreement with recent DFT studies [7,17,18].

DFT calculations show that, upon compression, a 2-L graphene film on SiC(0001) transforms into a diamond-like film with a Young's modulus of up to 1.01 TPa. The interface between the carbon film and substrate consists of a thin layer of bonds whose transverse mechanical strength is comparable to that of bulk SiC, whereas the surface of the diamond-like film consists of either chemically inert regions exposing for instance the well-known dimer C-C reconstruction (Fig. 4c) or regions with dangling bonds that may be passivated by H or -OH species [13]. This qualitative but important insight allows to model the process of a hard sphere indenting a hard ultra-thin film on a softer substrate. To this end, and to circumvent the limitations of continuum elasticity theory, we devise a classical scheme based on atomistic model structures and Hookean force fields. This simulation scheme, which is described in detail in the SI, is used to calculate the force vs. indentation depth curves for a substrate coated by a sub-nanometer-size film mimicking the elastic properties of either multilayer graphene or a diamond-like film, and thus exhibiting a transverse elastic modulus either 10 times smaller or 3 times larger than that of the substrate, respectively. As shown in Fig. 4d, these indentation simulations show that a SiC(0001) substrate coated by a stiff diamond-like film yields a force vs. indentation depth curve steeper than that of the bare SiC substrate, whereas a 5-L graphene film or a 2-L graphene film, which does not undergo any phase change, on SiC lead to a significant softening of the transverse mechanical response. These results are in good qualitative agreement with the experimental indentation curves reported in Fig. 1 and 2, demonstrating that only an ultra-stiff 2-L film, considerably stiffer than graphene, can lead to the stiffening effect, compared to bare SiC, reported in Fig. 2.

**Conclusions**

Sub-Å resolution indentation experiments, micro-hardness measurements, and conductive AFM investigations show that at room temperature, pressures of the order of 1-10 GPa can reversibly transform 2-L epitaxial graphene into a diamond-like ultra-stiff and ultra-hard phase. A SiC(0001) substrate coated with a two-layer epitaxial graphene film not only displays a stiffness similar to that of diamond but it also resists perforation by a diamond indenter at loads that can create plastic indents in bare SiC, one of the hardest materials known. All these properties are not observed for graphene films thicker than 3-5 layers, or consisting of just the single buffer layer, proving that 2-L graphene on SiC(0001) exhibits unique mechanical properties. DFT calculations show that the observed mechanical behavior of 2-L graphene stems from its ability to undergo a structural $sp^2$-to-$sp^3$ transformation. Such ability is hindered in 3-L or thicker graphene films by the occurrence of unfavorable layer stacking configurations, whereas in 2-L graphene, it is favored by the presence of a buckled buffer layer in contact with the reactive SiC(0001) substrate. DFT calculations show that diamond-like films obtained by compressing 2-L graphene may exhibit an elastic modulus as large as 1 TPa, and that formation of a diamond-like structure is not precluded by the lack of chemical species to passivate and stabilize the outer film surface. Our findings identify supported 2-L graphene as an interesting candidate for pressure-activated adaptive ultra-hard and ultra-thin coatings and for force-controlled dissipation switches. Our study opens up new ways to investigate graphite-to-diamond phase transitions at room temperature in low dimensional systems. Finally, this work suggests a new route to produce and pattern single layer diamond in graphene. For example, the diamond-like phase could be stabilized by combining local pressure with temperature, passivating gasses, or local heating [33], and applications could range from nanoelectronics to spintronics.

**Methods**

**Growth and characterization of epitaxial graphene films on SiC**

Epitaxial graphene films were grown on 4H-SiC on axis wafers by the confinement controlled sublimation method [20]. Details can be found in SI and in Refs. [20], [25] and in [34]. In Figures 1d and 1f of the manuscript, the buffer layer and 5-L graphene were grown on the (0001)-face (Si-face) of SiC. In Figure 1b, 10-L graphene was grown on the (000-1)-face (C-face) of 4H-SiC. In general, films with more than 10 layers were grown on the C-face. All the experiments reported here on the buffer layer and 2L graphene were performed on samples grown on the (0001)-face (Si-face) of SiC. Transmission Electron Microscopy (TEM) (Fig. 2 and Supplementary Fig. 1), Kelvin Probe Force Microscopy, Scanning TEM (Supplementary Fig. 2) and Raman spectroscopy measurements were performed to characterize the epitaxial graphene samples and probe the number of graphene layers. Details can be found in Ref. [35] and in the SI (see Supplementary Fig.s 1 and 2).

**MoNI stiffness measurements**

A detailed description of the MoNI technique is reported in the SI, as well as in Ref. [22] and in the Supplementary Information of Ref. [19]. Sub-Å vertical oscillations $\Delta z_{piezo}$ are applied at 1 kHz to the AFM tip via a piezoelectric stage rigidly attached to the AFM cantilever-tip system, and controlled by a Lock-in amplifier, while a constant normal force $F_z$ between the tip and the supported graphene films is maintained constant by the feedback loop of the AFM (see SI for more details). An important feature of MoNI is that the tip is first brought into contact with the sample and pressed until a force of the order of $10^2$ nN is reached; then the tip is retracted (in about ~10 s) from the surface while normal force variation $\Delta F_z$ at each fixed normal force $F_z$ are measured during the retraction. The integration of the measured $\Delta F_z/\Delta z_{piezo}$ at different normal loads $F_z$ allows us to acquire high-resolution force ($F_z$) indentation curves. MoNI experiments have been performed on standard samples such as ZnO single crystals and quartz to ensure the ability of MoNI to obtain reliable quantitative results. The typical error in force determination is $\Delta F_z = 0.5$ nN which translates into an error in the indentation position of about $\Delta z = 0.003$ nm, for the cantilever spring constant used in our study $k_{lev} = 173$ N/m. The estimated overall accuracy of the MoNI method is $\Delta z = 0.01$ nm. The MoNI experiments have been repeated with the same AFM tip back and forth from the SiC sample to the different graphene samples, to ensure that the tip remains unchanged. About 60 MoNI complete measurements, at different times, locations and for different samples have been acquired and show consistent results. More details are in the SI and Supplementary Fig. 4-5. Different methods (see SI and Fig. S3) are used to determine tip radius and cantilever spring constant, typical values are respectively ~100 nm and 173 N/m.

**Micro-hardness measurements**

The micro-hardness experiments presented in Fig. 3 were performed with an AFM sapphire cantilever having a diamond indenter attached to it (Micro Star TD10780, spring constant 151 N/m, and estimated tip radius 100 ± 50 nm). In a typical experiment, the sample is initially scanned with a set point force of 0.5-1.5 µN to record the initial topography and remove eventual organic residues from the surface [36]. The indentation force is applied at the center of the scanned region by directly approaching the surface with a set point force of 12 µN and an approaching speed of approximately 5 µm/s. After indentation, the surface is scanned again with a lower set point force (less than 0.5 µN) to assess the presence of the residual indent. Following [28], the hardness, $H$, is derived by using the following definition, $H = F/A$, where $F$ is the maximum force applied with the indenter and $A$ is the area of the residual indent determined by the analysis of the AFM topography image and using different equations as described in [28] and [29]. See more experiments with maximal loads up to 30 µN in Supplementary Fig. 6.

**Conductive AFM measurements**

Conductive AFM were performed on 2-L graphene and 5-L graphene films with AFM probes purchased from Bruker (model: SCM-PTSI). The tip is made of doped Si with Pt coating on the apex to enhance its conductivity. The spring constant of the cantilever is 2.8 N/m. Electric current flows between the tip and a silver-paste electrode deposited close to the edge of the sample (~300 μm). At each normal force (varying from 0 nN up to 350 nN), a 50 nm × 50 nm current AFM image was taken using a constant DC voltage of 3 mV. Then, the average current and standard deviation error are obtained from the current AFM images. More details are reported in SI and Supplementary Fig 7.

**Computational methods**

Density functional theory calculations were carried out by using tools in the QUANTUM-Espresso package [37]. We used norm-conserving pseudopotentials for all atomic species, a conventional parametrization for the generalized gradient approximation of the exchange and correlation energy [38], a plane-wave energy cutoffs larger than 80 Ry, and a semi-empirical corrective scheme to account for London dispersion interactions [39]. Details about supercells, model structures, DFT optimization calculations, and the method used to calculate the Young's modulus of both a diamond-like film and its interface with SiC are provided in the SI.

Indentation simulations relied on the use of atomistic model structures and Hookean force fields. We used cubic lattices with lattice constant of 2 Å to describe both substrate and 2D film. Substrate and 2D film are periodic in the xy-plane and finite in the perpendicular (z-) direction. In the xy-plane, substrate and 2D films include 40x40 cubic lattice sites, whereas in the perpendicular direction they include 15 layers for the substrate and 2 or 5 atomic layers to mimic 2D films with the elastic properties of 2-L or 5-L graphene, or a stiffer diamond-like film. The first (bottom) layer of the substrate is held fixed during the indentation simulations. Each lattice site is connected to its $1^{st}$, $2^{nd}$, and $3^{rd}$ nearest neighbor sites via a harmonic spring with a stiffness $k$. Sites in the same xy-plane are connected through springs with constant $k_{xy}$, whereas sites belonging to different layers are connected by springs with constant $k_z$. Values of these spring constants were calibrated to reproduce a Young's modulus of about 0.45 TPa for an isotropic substrate, a modulus of 1.2 TPa for an isotropic stiff material, and moduli parallel and perpendicular to the xy-planes of about 50 GPa and 1.2 TPa, respectively, for an anisotropic material used to mimic multilayer graphene films. Additional details can be found in the SI.

To describe a spherical indenter and its interaction with the lattice sites lattice sites of the substrate/2D film, we used a spherical Fermi-Dirac-type potential energy function, with parameters such as to mimic a tip with a radius of 5 nm and an elastic modulus close to that of diamond. Further details about the molecular-dynamics-based simulation scheme and the indentation curves shown in Fig. 4d are discussed in the SI and Supplementary Fig. 15.

## Acknowledgments

We acknowledge the support from the Office of Basic Energy Sciences of the US Department of Energy, and the European ERC 320796 MODPHYSFRICT. We also acknowledge the support by the CUNY High Performance Computing Center. We thank Dr. Tong Wang for his support with TEM measurements, and Prof. Michael Moseler for the useful discussions on indentation simulations.

## Author contributions

Y.G. and F.C. performed nanomechanics experiments and data analysis. T.C. carried out DFT calculations and indentation simulations. E.R. conceived and designed the experiments and analysed the data. A.B. and E.T analyzed the experimental data and delineate the modeling strategy. C.B., and W.d.H. synthesized the EG samples. All authors contributed to write the article.

## Additional information

Supplementary information is available in the online version of the paper. Reprints and permission information is available online at www.nature.com/reprints. Correspondence and requests for materials should be addressed to E.R. and. A.B.

## Competing financial interests

The authors declare no competing financial interests.

# References

1\. Cynn, H., Klepeis, J. E., Yoo, C. S. & Young, D. A. Osmium has the lowest experimentally determined compressibility. *Phys. Rev. Lett.* **88,** 135701 (2002).

2\. Narayan, J., Godbole, V. P. & White, C. W. Laser method for synthesis and processing of continuous diamond films on nondiamond substrates. *Science* **252,** 416-418 (1991).

3\. Jaglinski, T., Kochmann, D., Stone, D. & Lakes, R. S. Composite materials with viscoelastic stiffness greater than diamond. *Science* **315,** 620-622 (2007).

4\. Aust, R. B. & Drickamer, H. G. Carbon: A new crystalline phase. *Science* **140,** 817-819 (1963).

5\. Bundy, F. et al. The pressure-temperature phase and transformation diagram for carbon; updated through 1994. *Carbon* **34,** 141-153 (1996).

6\. Gorrini, F. et al. On the thermodynamic path enabling a room-temperature, laser-assisted graphite to nanodiamond transformation. *Sci. Rep.* **6** (2016).

7\. Horbatenko, Y. et al. Synergetic interplay between pressure and surface chemistry for the conversion of $sp^2$-bonded carbon layers into $sp^3$-bonded carbon films. *Carbon* **106,** 158-163 (2016).

8\. Khaliullin, R. Z., Eshet, H., Kühne, T. D., Behler, J. & Parrinello, M. Nucleation mechanism for the direct graphite-to-diamond phase transition. *Nat. Mater.* **10,** 693-697 (2011).

9\. Mao, W. L. et al. Bonding changes in compressed superhard graphite. *Science* **302,** 425-427 (2003).

10\. Odkhuu, D., Shin, D., Ruoff, R. S. & Park, N. Conversion of multilayer graphene into continuous ultrathin $sp^3$-bonded carbon films on metal surfaces. *Sci. Rep.* **3,** 3276 (2013).

11\. Scandolo, S., Bernasconi, M., Chiarotti, G. L., Focher, P. & Tosatti, E. Pressure-induced transformation path of graphite to diamond. *Phys. Rev. Lett.* **74,** 4015-4018 (1995).

12\. Xie, H., Yin, F., Yu, T., Wang, J.-T. & Liang, C. Mechanism for direct graphite-to-diamond phase transition. *Sci. Rep.* **4** (2014).

13\. Barboza, A. P. et al. Room-temperature compression-induced diamondization of few-layer graphene. *Adv. Mater.* **23,** 3014-3017 (2011).

14\. Rajasekaran, S., Abild-Pedersen, F., Ogasawara, H., Nilsson, A. & Kaya, S. Interlayer carbon bond formation induced by hydrogen adsorption in few-layer supported graphene. *Phys. Rev. Lett.* **111,** 085503 (2013).

15      Luo, Z. et al. Thickness-dependent reversible hydrogenation of graphene layers. *ACS Nano* **3,** 1781-1788 (2009).

16      Martins, L. G. P. et al. Raman evidence for pressure-induced formation of diamondene. *Nat. Commun.* **8,** 96 (2017).

17      Kvashnin, A. G., Chernozatonskii, L. A., Yakobson, B. I. & Sorokin, P. B. Phase diagram of quasi-two-dimensional carbon, from graphene to diamond. *Nano Lett.* **14,** 676-681 (2014).

18      Chernozatonskii, L. A., Sorokin, P. B., Kvashnin, A. G. & Kvashnin, D. G. Diamond-like $C_2H$ nanolayer, diamane: Simulation of the structure and properties. *JEPT Lett.* **90,** 134-138 (2009).

19      Gao, Y. et al. Elastic coupling between layers in two-dimensional materials. *Nat. Mater.* **14,** 714-720 (2015).

20      De Heer, W. A. et al. Large area and structured epitaxial graphene produced by confinement controlled sublimation of silicon carbide. *Proc. Natl. Acad. Sci.* **108,** 16900-16905 (2011).

21      Riedl, C., Coletti, C. & Starke, U. Structural and electronic properties of epitaxial graphene on SiC(0001): a review of growth, characterization, transfer doping and hydrogen intercalation. *J. Phys. D: Appl. Phys.* **43,** 374009 (2010).

22      Palaci, I. et al. Radial elasticity of multiwalled carbon nanotubes. *Phys. Rev. Lett.* **94,** 175502 (2005).

23      Lucas, M., Mai, W., Yang, R., Wang, Z. L. & Riedo, E. Aspect ratio dependence of the elastic properties of ZnO nanobelts. *Nano Lett.* **7,** 1314-1317 (2007).

24      Chiu, H. C., Kim, S., Klinke, C. & Riedo, E. Morphology dependence of radial elasticity in multiwalled boron nitride nanotubes. *Appl. Phys. Lett.* **101,** 103109 (2012).

25      Berger, C. et al. Electronic confinement and coherence in patterned epitaxial graphene. *Science* **312,** 1191-1196 (2006).

26      Kelly, B. T. *Physics of Graphite*. (Springer Netherlands, 1981).

27      Kumar, S. & Parks, D. M. Strain shielding from mechanically activated covalent bond formation during nanoindentation of graphene delays the onset of failure. *Nano Lett.* **15,** 1503-1510 (2015).

28      Richter, A., Ries, R., Smith, R., Henkel, M. & Wolf, B. Nanoindentation of diamond, graphite and fullerene films. *Diam. Relat. Mater.* **9,** 170-184 (2000).

29    Marcel Lucas, Ken Gall & Riedo, E. Tip size effects on atomic force microscopy nanoindentation of a gold single crystal. *J. Appl. Phys.* **104,** 113515 (2008).

30    Deng, X., Chawla, N., Chawla, K. K., Koopman, M. & Chu, J. P. Mechanical behavior of multilayered nanoscale metal-ceramic composites. *Adv. Eng. Mater.* **7,** 1099-1108 (2005).

31    Kulikovsky, V. et al. Hardness and elastic modulus of amorphous and nanocrystalline SiC and Si films. *Surf. Coat. Technol.* **202,** 1738-1745 (2008).

32    Kvashnin, A. G. & Sorokin, P. B. Lonsdaleite films with nanometer thickness. *J. Phys. Chem. Lett.* **5,** 541-548 (2014).

33    Wei, Z. et al. Nanoscale tunable reduction of graphene oxide for graphene electronics. *Science* **328,** 1373-1376 (2010).

34    Berger, C. et al. in *Graphene Growth on Semiconductors*   (eds N. Motta, F. Iacopi, & C. Coletti)    181-199 (Pan Stanford Publishing Pte. Ltd., 2016).

35    Filleter, T., Emtsev, K., Seyller, T. & Bennewitz, R. Local work function measurements of epitaxial graphene. *Appl. Phys. Lett.* **93,** 3117 (2008).

36    Gallagher, P. et al. Switchable friction enabled by nanoscale self-assembly on graphene. *Nat. Commun.* **7,** 10745 (2016).

37    Giannozzi, P. et al. QUANTUM ESPRESSO: a modular and open-source software project for quantum simulations of materials. *J. Phys. Condens. Matter* **21,** 395502 (2009).

38    Perdew, J. P., Burke, K. & Ernzerhof, M. Generalized gradient approximation made simple. *Phys. Rev. Lett.* **77,** 3865 (1996).

39    Grimme, S. Semiempirical GGA‐type density functional constructed with a long‐range dispersion correction. *J. Compu. Chem.* **27,** 1787-1799 (2006).

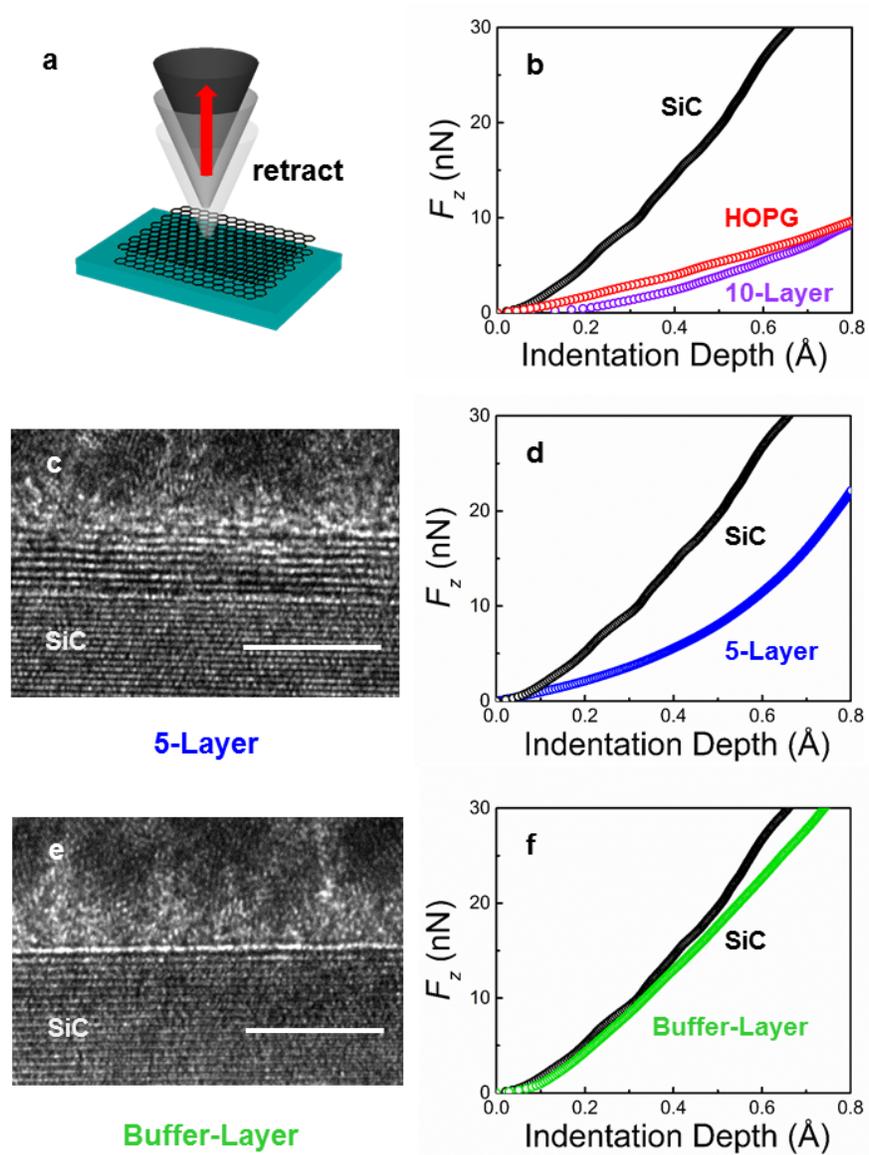

**Figure 1 | TEM images and experimental stiffness curves for multi-layer epitaxial graphene and buffer layer on SiC**. **a,** Schematics of the experiments performed in retracting mode. **b,** Indentation curves of SiC, HOPG and 10-L graphene on SiC (000-1). (**c**) TEM image and (**d**) indentation curve of 5-L graphene on SiC (0001). (**e**) TEM image and (**f**) indentation curve of the buffer-layer on SiC (0001). The MoNI curves in (**b**), (**d**) and (**f**) are taken with the same tip and they are averaged over 10 curves (see Supplementary Fig. 4 and 5), experimental errors are $\Delta F_z = 0.5$ nN and $\Delta z = 0.01$ nm. Scale bars in (**c**) and (**e**) are both 5 nm.

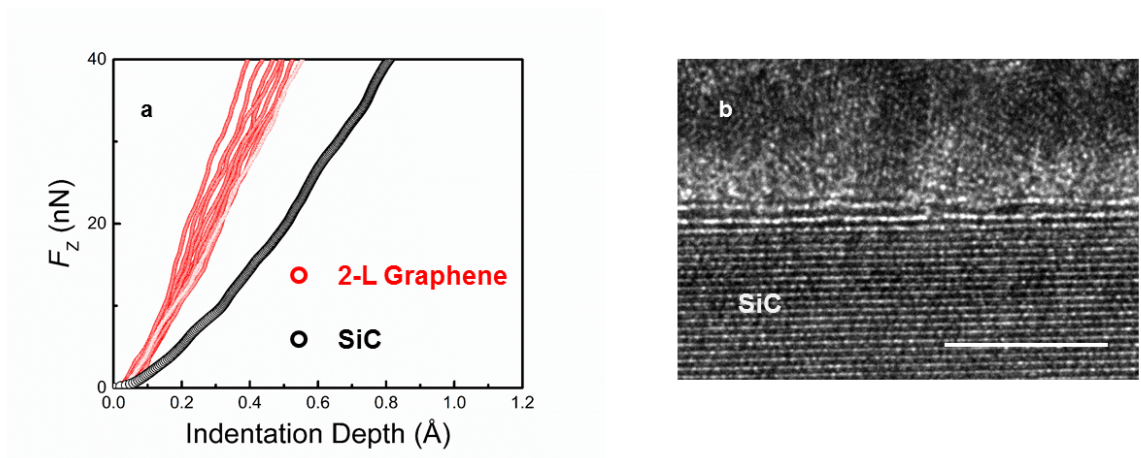

**Figure 2 | Experimental stiffness measurements in 2-L graphene.** Experiments showing a new ultra-stiff phase in two-layer graphene at room temperature upon indentation. **a,** Experimental indentation curves in two-layer epitaxial Graphene (red), and SiC (black). For 2L graphene, the curves were acquired in different positions on different samples. **b,** TEM image of two-layer epitaxial graphene on SiC. The scale bar is 5 nm.

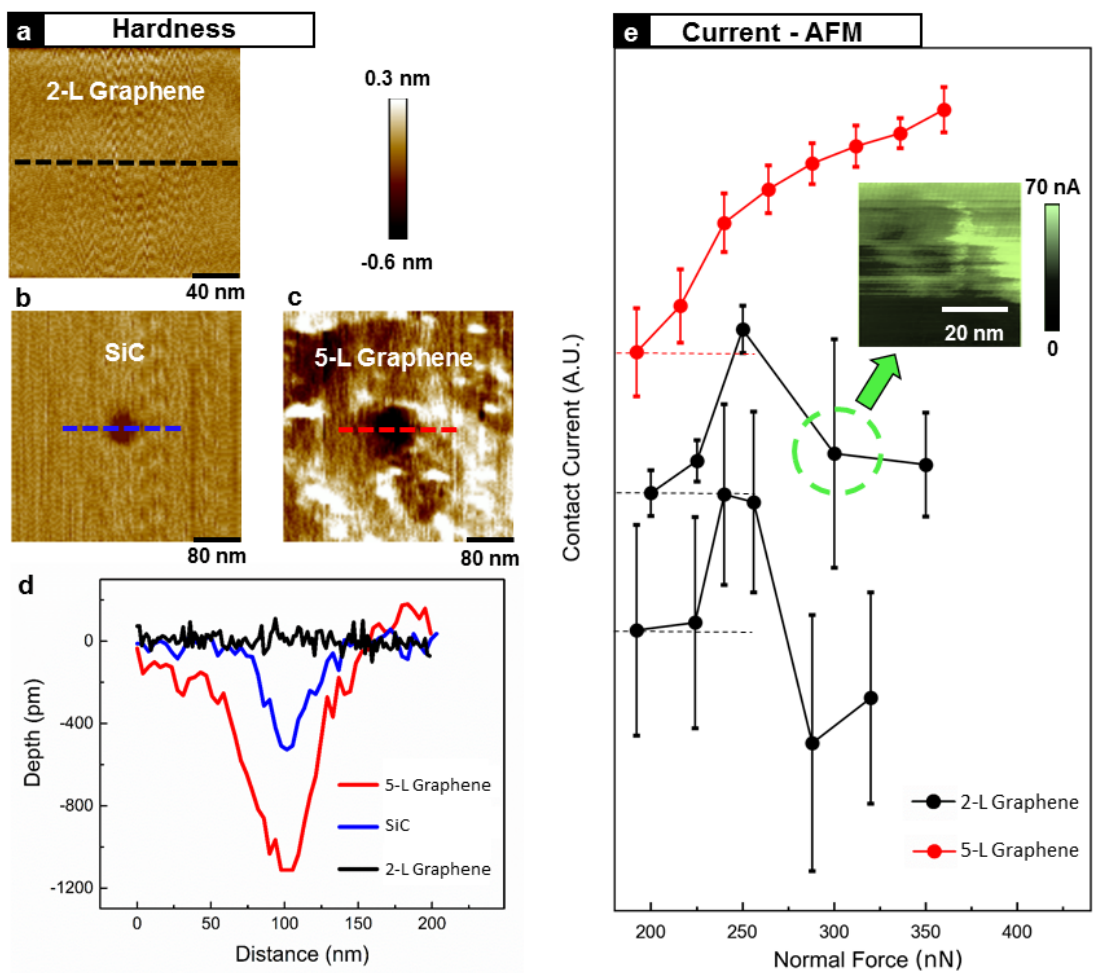

**Figure 3 | Micro-hardness and Current-AFM measurements**. AFM topographical images of the residual indents in 2-L graphene (**a**), SiC (**b**), and 5-L graphene (**c**) upon indentation with a diamond indenter. **d**, Cross section profile of the residual indents in 2-L graphene, SiC and 5-L graphene as shown in the AFM topographic images in (**a-c**), respectively. **e**, Three stacked curves of the average current signal vs. normal load as measured by C-AFM (see Supplementary Fig. 7) on areas of 50 x 50 nm$^2$ on two different 2-L graphene films (black curves) and a 5-L graphene film (red curve). A current drop is observed at ~ 260 nN only for 2-L graphene samples. The inset shows a current image recorded on 2-L graphene at 300 nN while scanning from top to bottom, indicating a drop in current occurring after a few scans. These current vs. load curves are reversible when decreasing the load.

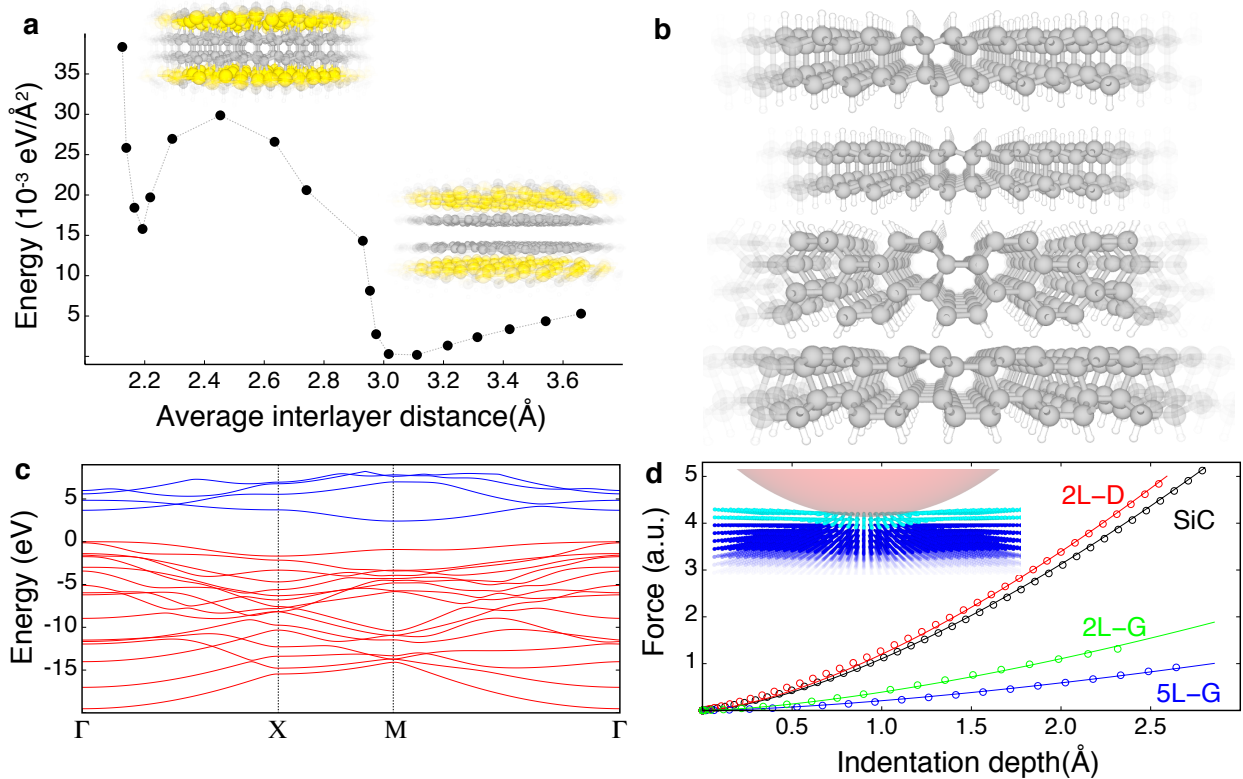

**Figure 4 | DFT and Indentation Calculations. a,** Energy per unit area obtained by DFT calculations for a two-layer graphene film sandwiched between mirroring H-CSi layers; the graphene layers include 4×4 elementary unit cells in contact with a SiC(0001) substrate having a periodicity of $2\sqrt{3} \times 2\sqrt{3} R30^0$. Energy values are referred to the smallest ones of the set and are plotted vs. the average distance between the carbon layers of graphene (right inset) and diamond-like (left inset) films. **b,** Diamond-like films with surfaces exposing chair (top two image) and boat (bottom two images) conformations, obtained by compacting two graphene layers with, from the top to the bottom, a AB, AA, AB, and $A_BC$ stacking configuration. The top three images show films with both surfaces passivated by H. The image at the bottom of this panel has H atoms passivating only one of the two equivalent surfaces; the electronic band structure of this latter model is shown in panel **c**, with valence and conduction bands shown in red and blue colors, respectively. **d,** Force vs. indentation curves obtained by using a classical scheme based on atomistic model structures and Hookean force fields. Circles show calculated data, whereas solid lines show fits using a Hertz law. Simulations were carried out for a bare substrate (SiC, black color), and for the same substrate coated with a 2-L film with the elastic moduli of a diamond-like (2-L D, red color) or graphene (2-L G, green color) film, as well as for a substrate coated with a 5-L graphene film (5-L G, blue color). The inset illustrates a spherical indenter (pale red) pressing on a substrate material (blue) coated with thin film of a stiffer material (cyan).